\documentstyle[aps,floats,twocolumn,epsf]{revtex}
\tighten
\begin{document}
\newlength{\tdwidth}
\tdwidth \columnwidth
\newlength{\tdtopskip}
\tdtopskip \topskip
\draft
\title{Enhanced Electron-Phonon Coupling and its
Irrelevance to High T$_{c}$ Superconductivity}
\author{$^{1}$T. P. Devereaux, $^{2}$A. Virosztek, $^{2}$A. Zawadowski,
$^{3}$M. Opel, $^{3}$P. F. M\"uller, $^{3}$C. Hoffmann, $^{3}$R. Philipp, 
\linebreak
$^{3}$R. Nemetschek, $^{3}$R. Hackl, $^{4}$A. Erb, $^{4}$E. Walker, 
$^{5}$H. Berger and $^{5}$L. Forr\'o}
\address{$^{1}$Department of Physics, George Washington University, Washington,
DC 20052}
\address{$^{2}$Institute of Physics and Research Group of the Hungarian Academy
of Sciences, Technical University of Budapest, H-1521
Budapest, Hungary and Research Institute for Solid State Physics, 
P.O.Box 49, H-1525 Budapest, Hungary}
\address{$^{3}$Walther Meissner Institut, Bayerische Akademie der
Wissenschaften, Garching, D-85748, Germany}
\address{$^{4}$DPMC, Universit\'e de Gen\`eve, CH-1211 Gen\`eve, Switzerland}
\address{$^{5}$EPFL, Ecublens, CH-1015 Lausanne, Switzerland}
\address{~\parbox{14cm}{\rm 
\medskip
It is argued that the origin of the
buckling of the CuO$_{2}$ planes in certain cuprates 
as well as the strong electron-phonon coupling of the $B_{1g}$
phonon is due to the electric field across the planes induced by
atoms with different valence above and 
below. The magnitude of the electric field is deduced from new 
Raman results on YBa$_{2}$Cu$_{3}$O$_{6+x}$ and 
Bi$_{2}$Sr$_{2}$(Ca$_{1-x}$Y$_{x}$)Cu$_{2}$O$_{8}$ with different O and 
Y doping, respectively. In the latter case it is shown that the 
symmetry breaking by replacing Ca partially by Y enhances
the coupling by an order of magnitude, while the superconducting $T_c$
drops to about two third of its original value.
\vskip0.05cm\medskip 
}}
\maketitle
\narrowtext

Recently there have been several suggestions\cite{Sug} that out of plane
oxygen vibrations of the CuO$_2$ plane may play an important role in
mediating superconducting pairing
in certain high
temperature superconductors. It is also known that the presence of buckling
(when the oxygen atoms are placed outside of
the plane of the copper atoms) 
correlates with
a strong interaction of a particular out-of-phase 
c-axis vibration (the so-called $B_{1g}$ phonon)
with planar quasiparticles\cite{alt}. As such,
considerable attention has been focused on the strong interaction of
electrons with this particular phonon and its connection to
phonon-mediated $d_{x^{2}-y^{2}}$ superconductivity\cite{SA,djs}.

The present paper provides the first direct experimental evidence to
support that the buckling and consequent large value of the electron-
phonon interaction are indeed correlated as a result of a local
crystal electric field surrounding the vibrating atoms. 
Using the above ideas in conjunction with a three band model for the
electron system, the strength of the electron phonon coupling and
consequently the value of the electric field perpendicular to the plane
can be determined by fitting the Fano interference in the $B_{1g}$ Raman
spectra. We find that
even though the electron-phonon coupling can be engineered in
different materials to increase
by an order of magnitude, the effect has essentially no correlation
with T$_{c}$ of the material. Therefore we argue that the $B_{1g}$
phonon is irrelevant to superconductivity in the cuprates.


Raman scattering
data are presented in the normal state for different doping levels of 
YBa$_{2}$Cu$_{3}$O$_{6+x}$ (Y-123) 
and Bi$_{2}$Sr$_{2}$CaCu$_{2}$O$_{8}$ (Bi-2212). 
Experimentally the
electron-phonon coupling can be determined from the line shape of the
$B_{1g}$ Raman phonon for both Y-123 where the
surroundings of the CuO$_2$ plane is highly asymmetric and Bi-2212
where the plane is placed in a more symmetrical environment with respect
to the charges of the nearby atoms.

The resulting electron-phonon coupling is then used to interpret
the Fano interference in the Raman spectra of Y-123, where
the light is scattered by both the $B_{1g}$ phonon and the electronic
charge fluctuation in the plane described above. 

However, for Bi-2212 the $B_{1g}$
phonon has an almost Lorentzian line shape indicating the lack of any
substantial coupling between that phonon and the charge transfer between
the two oxygens in accordance with the absence of a significant 
electric field.
This striking difference in the data taken on these two groups of materials
can be considered as strong evidence in favor of the linear (involving
one phonon) electron-phonon coupling due to the electric field. 
In order to complete this argument, the data for a Bi-2212 sample
with the calcium partially (38\%) substituted by yttrium are presented,
where the doping most likely breaks locally the reflection symmetry
through the CuO$_2$ plane, thereby making this material similar to the
asymmetrical ones.

\begin{figure}
\vskip  0cm
\epsfxsize=8cm
\epsfysize=4cm
\epsffile{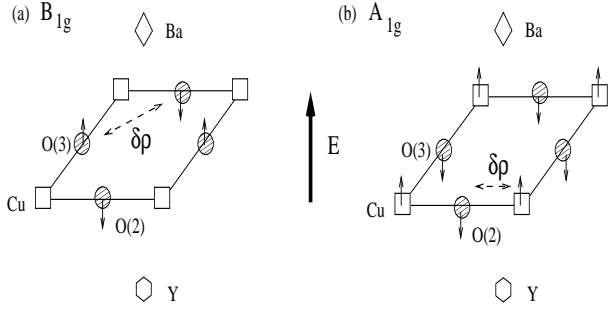}
\vskip 0.3cm
\caption{The unit cell of the CuO$_{2}$ plane shown with the atomic 
displacements corresponding to the $B_{1g}$ (a) and $A_{1g}$ (b)
phonons. The electric field $E$ perpendicular to the planes is due
to the asymmetric environment (Ba$^{2+}$ above, Y$^{3+}$ below), and 
causes a static
deformation of the $A_{1g}$ type, usually referred to as buckling.
$\delta\rho$ and the double arrows denote the charge transfer which
accompanies the lattice vibration.}
\label{fig.1}
\end{figure}

Finally, the value obtained for the electron-phonon coupling
is in excellent agreement with
the value calculated using the electric field value determined by the
experimentally observed buckling\cite{Jorg,further} and the restoring force 
calculated from the frequency of the $A_{1g}$ phonon.
The change in the doping is essential in light of the closeness of the
Fermi level to a van Hove singularity in the density of states.

Much of the development of the crystal field model was given in
Ref. \cite{DVZ1}. However, the electronic dispersion is now calculated
for a three band model including direct $O-O$ hopping $t^{\prime}$
as well as $Cu-O$ hopping $t$ \cite{further}.
As in Ref. \cite{DVZ1} we consider only a reduced one band model
appropriate for near half filling and take only the upper band 
into account.
The electron - phonon reduced Hamiltonian
for the $B_{1g}~~ {\bf q}=0$ phonon was given as
\begin{equation}
H_{el-ph}={1\over{\sqrt{N}}}\sum_{{\bf k},\sigma} g({\bf k}) 
d^{\dagger}_{{\bf k},\sigma}d_{{\bf k},\sigma}[c+c^{\dagger}],
\label{one}
\end{equation}
where $d_{{\bf k},\sigma}$ annihilates an electron of spin $\sigma$ and
momentum ${\bf k}$, and $c^{\dagger}$ creates a $B_{1g}$ phonon mode of
wavevector ${\bf q}=0$. The 
coupling constant $g({\bf k})$ 
of the $B_{1g}$ mode to an electron with momentum ${\bf k}$ 
was evaluated in Ref. \cite{DVZ1} and is given by
\begin{equation}
g({\bf k})= eE\sqrt{{\hbar\over{2M\omega_{B_{1g}}}}} {1\over{\sqrt{2}}}
[\mid\phi_{x}({\bf k})\mid^{2}-\mid\phi_{y}({\bf k})\mid^{2}],
\label{two}
\end{equation}
where $M$ is the oxygen mass, $\omega_{B_{1g}}$ is the phonon frequency,
and $E$ is the ${\bf \hat z}$- component of the electric crystal field
at both the $O(2)$ and $O(3)$ sites.
The functions $\phi_{x,y}$ are the amplitudes of the $O-$orbitals in
the wave functions from the three band model\cite{further}.

In Ref. \cite{DVZ1} 
a generalized form of the Breit-Wigner or Fano lineshape
describing the interaction of a discrete excitation (phonon) with an
electronic continuum was given in terms of the channel
dependent electronic susceptibility $\chi_{\lambda}$\cite{nfl}, 
the electron-phonon coupling constant,
the effective photon-phonon coupling constant $g_{p-p}$, and the intrinsic 
damping $\Gamma _{\lambda}^{i}$ of the
phonon lineshape 
due to e.g. an anharmonic lattice potential. 
The expression for the full Raman response measured
in channel $\lambda$ was given as
\begin{eqnarray}
&&\chi_{\lambda,full}^{\prime\prime}(\omega)=
{(\omega+\omega_{a})^{2}\over{(\omega^{2}-\hat\omega_{\lambda}^{2})^{2}+
[2\omega_{\lambda}\Gamma_{\lambda}(\omega)]^{2}}}\nonumber\\
&&
\times\Bigg\{\gamma_{\lambda}^{2}\chi_{\lambda}^{\prime\prime}(\omega)\left[
(\omega-\omega_{a})^{2}+4\Gamma_{\lambda}^{i}\Gamma_{\lambda}(\omega)
\left({\omega_{\lambda}\over{\omega+\omega_{a}}}\right)^{2}\right]
\nonumber \\
& &+4g_{p-p}^{2}\Gamma_{\lambda}^{i}
\left({\omega_{\lambda}\over{\omega+\omega_{a}}}\right)^{2}
[1+\lambda(\omega)/\beta]^{2}\Bigg\}.
\label{three}
\end{eqnarray}

The parameters
are chosen to model the background spectrum seen of the normal state 
measured via Raman scattering. Then the remaining parameters are chosen
to fit the Fano profile. Specifically, the effect of the parameters is
as follows: the photon-phonon coupling constant $g_{p-p}$
determines the position of
the antiresonance $\omega_{a}$
of the Fano profile, the electron-phonon coupling constant $g$
determines the asymmetry of the lineshape around the phonon position
$\hat\omega$, which differs from its position $\omega$ in the absence of
electron-phonon coupling, and the photon-electron coupling constant
$\gamma$ determines the overall strength of the background continuum
under the phonon measured via Raman scattering\cite{param}. 
Here $\gamma_{B_{1g}}$ is the projected part of the electron-photon
vertex $\gamma({\bf k})$ which possesses $B_{1g}$ symmetry. 
$g_{B_{1g}}^{2}$ is the average of the coupling $\mid g({\bf k})\mid^{2}$
over the Fermi surface including the electron density for the two
spins at a temperature $T$. We can now evaluate the Fano lineshape as a 
function of doping.

Before a comparison of the theoretical predictions with the data is made
a few experimental details are given and the results are summarized.
The Raman experiments were performed in back-scattering 
geometry, with the resolution set at 8 cm$^{-1}$
being completely sufficient for the observation of the changes.
The coordinate system is locked 
to the $Cu-O$ bonds with $x$ = [100], $x^{\prime}$ = [110], etc. 
All symmetries refer to a tetragonal point group. The $B_{1g}$ 
phonon and continuum are projected out with $x^{\prime}y^{\prime}$ 
polarization.
The samples we used were of superior quality which manifests itself in a
small transition width to the superconducting state and a large
intensity ratio $I_{phonon}$ to $I_{continuum}$. Y-123 in particular
was prepared in BaZrO$_3$ resulting in an unprecedented purity of
99.995\%\cite{Erb}.
More details of the samples will be described in 
another publication.

Results obtained at $B_{1g}$ symmetry for Y-123 are 
plotted in Fig. 2 (a)-(c). All spectra are divided by the 
Bose-Einstein thermal function in order to get the 
response $\chi^{\prime\prime}_{\lambda,full}$ 
as described in Eq. (\ref{three}). As a result of 
doping the shape of the $B_{1g}$ phonon at approximately 
330 cm$^{-1}$ changes considerably. At low doping it is 
narrow and close to a Lorentzian. When carriers are 
added the line broadens and becomes more asymmetric 
exhibiting a Fano-type dependence on frequency.
$I_{phonon}/I_{continuum}$ decreases since  
the $B_{1g}$ continuum gains intensity\cite{katsu}. 
\begin{figure*}
\epsfxsize=18cm
\epsfysize=5cm
\epsffile{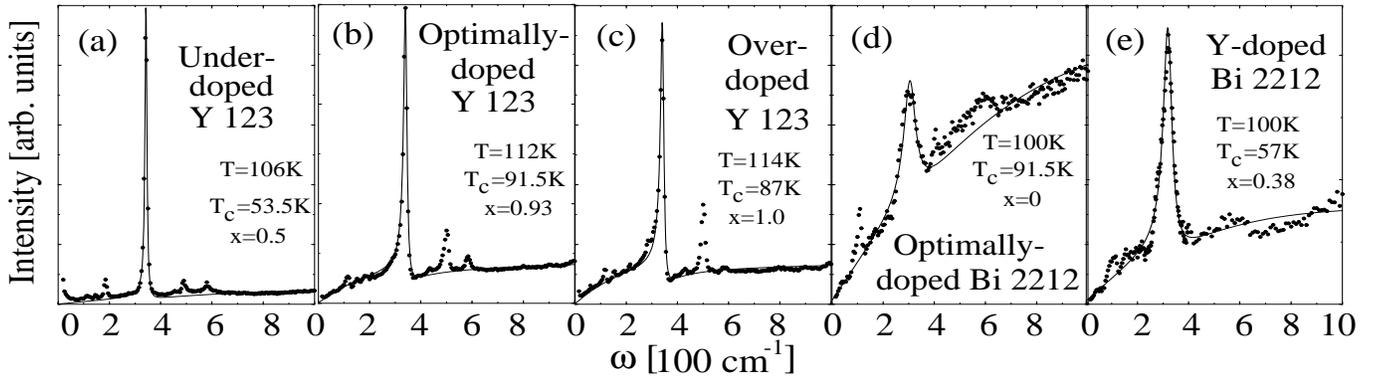}
\vskip 0.4cm
\columnwidth \textwidth
\caption{Comparison of the theory from the crystal field model to the 
$B_{1g}$ Raman results on (from left to right) under-doped,
optimally-doped, over-doped YBa$_{2}$Cu$_{3}$O$_{6+x}$, respectively,
and optimally-doped (second panel from right) and $Y-$doped
Bi$_{2}$Sr$_{2}$(Ca$_{1-x}$Y$_{x}$)Cu$_{2}$O$_{8+\delta}$ (right panel), 
respectively, and the scales for each are arbitrary.
The parameters used are listed in Table I.}
\columnwidth \tdwidth
\label{fig.2}
\end{figure*}
Excellent fits to the data on optimally-doped Y-123
in both the superconducting and normal
states were given in \cite{DVZ1}, which, when interpreted, supported
evidence for 
$d_{x^{2}-y^{2}}$-pairing in this material. We now
extend these results by examining fits to the normal state data of
different samples and dopings. We find that once again excellent
fits to the data can be obtained (Fig. 2). The respective fitting parameters
for $\chi_{\lambda}$ (see \cite{nfl}) are given in Table I.

We now can make a comparison to the crystal field model predictions for the
electron-phonon coupling constant. We see from the fits that for Y-123 
the coupling increases up to a
factor of two with increasing doping. This can be understood in part by
an increase in the density of states at the Fermi level with an increase in
oxygen doping as the van Hove singularity moves closer to the Fermi level.
Our calculations yield the following
values for $\lambda$ as a function of filling factor $<n>$: 
for $<n>=0.875, \lambda=0.0321$, for $<n>=0.85, \lambda=0.0382$,
and for $<n>=0.8, \lambda=0.0482$\cite{further}.
Here we have taken $T=100K, t=1.6eV, t^{\prime}/t=0.35,$ 
the difference of $Cu$ and $O$ site energies 
$\varepsilon=1eV, \omega_{B_{1g}}=348$cm$^{-1}$,
$M=16 m_{P}$, where $m_{P}$ is the proton mass, and used a value of
the electric field which is $1.3$V/\AA\cite{DVZ1,further}. 
We observe that these values of $\lambda$
are quite close to the values used to obtain the fits in Fig. 2. Given
the assumption that the hopping and site energy parameters\cite{DVZ1}
are independent of doping, the agreement is quite good. The agreement could
be refined once parameter choices for various levels of doping are
determined via, e.g., fitting to experimentally observed Fermi surfaces.

In Bi-2212 (Fig. 2 (d)) the $B_{1g}$ 
phonon line is much weaker while the continuum has 
roughly the same cross section as comparable Y-123 
(Fig. 2 (b)). The bigger line width comes at least in part 
from inhomogeneous broadening as Bi-2212 single 
crystals cannot be synthesized stoichiometrically in 
equilibrium\cite{revaz}. The electron-phonon coupling for Bi-2212 
obtained from the fit is more
than one order of magnitude smaller than for any of the Y-123 compounds
and as a result the weight of the phonon is also reduced by at least
the same amount.
Indeed, in our model we would expect that the electron-phonon coupling
is significantly smaller
since the crystal electric field must be much weaker (the charges above and
below the planes are not so drastically different as in Y-123)
in this compound. We attribute
then the small electron-phonon coupling 
to a much smaller local asymmetry and also to
local inhomogeneities which manifest themselves in the larger 
linewidth of the phonon. A weaker spontaneous symmetry breaking is
also possible.

Inhomogeneities can be introduced in a controlled way by replacing
the Ca with 
(Ca$_{0.62}$Y$_{0.38}$) in Bi-2212. It is
expected that significant parts of the CuO$_2$ planes experience a 
stronger local
field loosing the reflection symmetry through that plane. Most likely,
the Y replaces big regions of the Ca planes, just like in Y-123 where
upon oxygen doping of the chains some of the chains are more doped than
the others\cite{Janossy}.

To test this idea we look at the measurements of Bi-2212 which
has been doped with Y in place of Ca.
When Y is 
doped in for Ca the line gains intensity and shifts by 
15-20 cm$^{-1}$ (Fig. 2 (e)). A reminder of the line 
found in Y-free crystals is still seen as a shoulder on the 
left-hand side of the
new line. Since here the valence of the 
Y (+3) is different
from that of Ca (+2), once again the mirror plane symmetry is broken and
we would expect a much larger electron-phonon coupling than in the
undoped compound. The fit to the $B_{1g}$ data
using Eq. (\ref{three}) is given in Fig. (2e). The coupling constant
(see Table I.)
is indeed increased over that of the undoped compound when Y is introduced,
and in accordance the intensity of the phonon line is also essentially
enhanced. The fact that the Fano effect increases
gives strong support for the crystal field coupling model as the
driving source of electron-phonon coupling.

The electric field E perpendicular to the CuO$_{2}$ plane 
results in an $A_{1g}$-type static distortion of the plane since 
the charges of oxygen and copper are different. 
For an estimation of the buckling\cite{Jorg}, a simplified 
model is sufficient\cite{further}
where Cu is pinned rigidly to the 
elementary cell, and the oxygen moves in a harmonic 
potential being characterized by the frequency of the 
$A_{1g}$ phonon at $\omega_{A_{1g}}=435$ cm$^{-1}$\cite{frozen}. 
The restoring 
force at the buckling amplitude $\Delta z$ must balance the 
electric force acting on the oxygen with charge $q = 
- 1.75 e$. Thus, $qE = M\omega_{A_{1g}}^2\Delta z$ 
must hold. With the experimental value $\Delta z = 0.24 \AA$\cite{Jorg}, 
$E = 1.53 V/\AA$ 
is obtained which is close both to the $1.3 V/\AA$ used for 
estimating the electron-phonon coupling strength and 
to the theoretically calculated number\cite{ladik}. On the other 
hand, in the case of CuO$_{2}$ planes in a more symmetrical 
environment as in Bi-2212 and the infinite-layer 
compound CaCuO$_{2}$ the buckling, if it exists at all, is at 
least an order of magnitude smaller as found in 
structural studies\cite{Karp}. 

\topskip \tdtopskip
In summary, studying Y-123 with different doping levels 
we have shown that the electric field across
the CuO$_{2}$ planes is sufficiently 
strong to produce both the observed buckling and the 
strong electron-phonon coupling for 
the Fano line shape. In 
order to check this idea experiments were performed on 
Bi-2212 with and without Y doping. While the sample 
without Y shows very weak electron-phonon coupling, 
the interaction is enhanced by an order of magnitude 
and becomes comparable to the one in Y-123 if the local 
reflection symmetry is broken more substantially
by replacing part of the Ca by Y. 
As doping Bi-2212 with Y results in a change of
T$_{c}$ from T$_{c}=91.5K$ to T$_{c}=57K$ along with
a large increase of the coupling $\lambda$, the $B_{1g}$
phonon can not play an important role
for the superconductivity, in agreement with the conclusion
of Savrasov and Andersen\cite{SA}.

\begin{table}
\begin{tabular}{|l|l|l|l|l|l|l|l|l|}
{Fig.}&{$\alpha$}&{$\hat\omega$}& 
{$\omega$}&{$\tau^{*-1}$}& 
{$\omega_{a}$}&{$\Gamma_{i}$}&  
{$\lambda$}&{scale}\\\hline 
(2a)&0.55&343&347.5&3000&352.5&4&0.0257&22\\ \hline 
(2b)&0.95&340.5&348&1200&349.3&6.5&0.0426&20\\ \hline 
(2c)&0.75&342&352&900&352.5&6&0.056&30\\ \hline 
(2d)&0.1&305&305.2&1600&306.5&28&0.00131&20\\ \hline 
(2e)&0.2&320&322&1200&326&20&0.0124&10\\
\end{tabular}
\vskip 0.5cm
\caption{Summary of fitting parameters used in Fig. (2). All quantities
except $\alpha, \lambda,$ and the scale factor (dimensionless)
are given in units of cm$^{-1}$. Also
$\omega_{c}=12,000$~cm$^{-1}$ and $\beta^{\prime}=3.3$ have been used.
A scale factor was multiplied to Eq. (\ref{three}) to account
for the overall magnitude of the cross section. Since the experimentally
determined response is in arbitrary units, this has no affect on
the conclusions.}
\end{table}

This work was supported by the Hungarian National Research Fund under
Grant Nos. OTKA T020030, T016740, T02228/1996, T024005/1997. 
Acknowledgment (T.P.D.) is made
to the Donors of The Petroleum Research Fund, administered by the 
American Chemical Society, for partial support of this research.
A. Z. is grateful for enlightening discussions with O. K. Andersen and
W. Pickett and for
the support by the Humboldt Foundation. 
The experimental work was supported by the Bayerische Forschungstiftung via
the consortium FORSUPRA. We are
grateful to the BMBF for financial support via the program
``Bilaterale wissenschaftlich-technische Zusammenarbeit'' under grant no. 
WTZ-UNG-052-96. One of us (T.P.D.) was partially supported by the
American Hungarian Joint Fund No. 587.


\begin{references}
\bibitem{further} Further details will appear in a forthcoming
publication.
\bibitem{param}
These parameters are related as in \cite{DVZ1}
$$\beta=2g_{p-p}g_{B_{1g}}[\gamma_{B_{1g}}\omega_{B_{1g}}]^{-1},~~~~~~~~
\omega_{a}^{2}=\omega_{B_{1g}}^{2}[1+\beta],$$
$$\lambda(\omega_{B_{1g}})=g_{B_{1g}}^{2}/\omega_{B_{1g}},$$
$$\Gamma_{B_{1g}}(\omega_{B_{1g}})=\Gamma^i_{B_{1g}}+
g_{B_{1g}}^{2}
\chi^{\prime\prime}(\omega_{B_{1g}})/N_{F},$$ 
with $N_{F}$ the density of states per spin at the Fermi level.
\end{references}
\end{document}